\documentclass[%
,aps%
,twocolumn%
,secnumarabic%
,amssymb, amsmath,nobibnotes, aps, prl, floatfix，preprint,groupedaddress,superscriptadress]{revtex4-2}
\usepackage{amsmath,epsfig,amssymb,subfigure,bm,dsfont}
\usepackage{lipsum}

\begin{document}

\title{Simulating a Chern Insulator with $C=\pm2$ on Synthetic Floquet Lattice}

\author{Lingxiao~Lei$^{1}$, Weichen~Wang$^{2}$}
\author{Guangyao~Huang$^{2}$}
\email{guangyaohuang@quanta.org.cn}
\author{Shun~Hu$^{2}$, Xi~Cao$^{3}$, Xinfang~Zhang$^{2}$}
\author{Mingtang~Deng$^{2,4}$}
\email{mtdeng@nudt.edu.cn}
\author{Ping-Xing~Chen$^{1,4}$}
\email{pxchen@nudt.edu.cn}

\affiliation{$^{1}$Institute for Quantum Science and Technology, College of Science, National University of Defense Technology, Changsha 410073, China\\
$^{2}$Institute for Quantum Information \& State Key Laboratory of High Performance Computing, \\ College of Computer Science and Technology, National University of Defense Technology, Changsha 410073, China\\
$^{3}$Greatwall Quantum Laboratory, Changsha 410006, China\\
$^{4}$Hefei National Laboratory, Hefei 230088, China}

\date{\today}

\begin{abstract}
The synthetic Floquet lattice, generated by multiple strong drives with mutually incommensurate frequencies, provides a powerful platform for the quantum simulation of topological phenomena. In this study, we propose a 4-band tight-binding model of the Chern insulator with a Chern number $C=\pm2$ by coupling two layers of the half Bernevig-Hughes-Zhang lattice and subsequently mapping it onto the Floquet lattice to simulate its topological properties. To determine the Chern number of our Floquet-version model, we extend the energy pumping method proposed by Martin \textit{et al.} [\href{https://doi.org/10.1103/PhysRevX.7.041008}{Phys. Rev. X \textbf{7}, 041008 (2017)}] and the topological oscillation method introduced by Boyers \textit{et al.} [\href{https://doi.org/10.1103/PhysRevLett.125.160505}{Phys. Rev. Lett. \textbf{125}, 160505 (2020)}], followed by numerical simulations for both methodologies. The simulation results demonstrate the successful extraction of the Chern number using either of these methods, providing an excellent prediction of the phase diagram that closely aligns with the theoretical one derived from the original bilayer half Bernevig-Hughes-Zhang model. Finally, we briefly discuss a potential experimental implementation for our model. Our work demonstrates significant potential for simulating complex topological matter using quantum computing platforms, thereby paving the way for constructing a more universal simulator for non-interacting topological quantum states and advancing our understanding of these intriguing phenomena.
\end{abstract}

\maketitle

{\it Introduction.} Recently, the concept of synthetic dimensions has attracted considerable attention in simulating topological quantum phenomena using controllable quantum systems owing to their enhanced flexibility compared to that of real materials~\cite{Ozawa_Nat.Rev.Phys._2019}. For instance, by employing hyperfine states~\cite{Celi_Phys.Rev.Lett._2014,Stuhl_Science_2015,Mancini_Science_2015} and spatial eigenmodes~\cite{Price_Phys.Rev.A_2017,Lustig_Nature_2019,Wang_Phys.Rev.Lett._2024} in atom/ion systems as well as orbital angular momentum~\cite{Luo_NC_2015,Cardano_NC_2017,Wang_Phys.Rev.Lett._2018} and Frequency modes~\cite{T.Ozawa_PhysRevA_2016, L.Yuan_OpticsLetters_2016} in photonic systems as synthetic dimensions, a wide range of topological quantum states have been successfully simulated. Among these approaches, the (synthetic) Floquet lattice~\cite{Malz_Phys.Rev.Lett._2021,Zhong_Phys.Rev.A_2023}, also referred to as the Floquet frequency lattice~\cite{Yu_Phys.Rev.Lett._2023}, exhibits universality across various experimental platforms. The Floquet lattice is generated by multiple incommensurate frequencies~\cite{Martin_Phys.Rev.X_2017} and provides synthetic dimensions whose quantity is determined by the number of incommensurate drives.

The synthetic Floquet lattice was initially reported in Ref.~\onlinecite{Martin_Phys.Rev.X_2017}, where Martin et al. investigated the half Bernevig-Hughes-Zhang (half-BHZ) model~\cite{Liu_Phys.Rev.Lett._2008,Qi_Phys.Rev.B_2006} constructed on the Floquet lattice space. In order to determine the Chern number of the Floquet-version half-BHZ model, an energy pumping method was proposed~\cite{Martin_Phys.Rev.X_2017} and experimentally realized in a superconducting quantum computing system~\cite{Malz_Phys.Rev.Lett._2021}. The Floquet half-BHZ model has also been simulated using a nitrogen-vacancy center quantum system~\cite{Boyers_Phys.Rev.Lett._2020}, where another Chern number measuring technique known as topological oscillation is employed. 

Despite the remarkable achievements made in the aforementioned research, these studies only considered the simplest half-BHZ model with a Chern number $C=\pm1$, which cannot capture the $\mathbb{Z}$ invariant nature of Chern number\cite{Chiu_RMP_2016}. Moreover, there has been a growing interest in exploring Chern insulators with higher Chern numbers across various fields. For instance, previous studies have reported the observation of Chern numbers $C=2,3,4$ in photonic crystals~\cite{Skirlo_PRL_2015}, as well as the successful engineering of topological phases with enhanced Chern numbers in ultracold atom gases~\cite{Alase_PRA_2021,Mateusz_SciPP_2021}. Furthermore, A recent study suggest that a monolayer of ${\rm Ti_2SiCO_2}$, when driven by circularly polarized light, exhibits characteristics of a Chern insulator with $C=\pm 2$~\cite{Liu_PRB_2024}.

In this study, we propose a 4-band tight-binding model (bilayer half-BHZ model) with an enhanced Chern number ($C=\pm2$), which is subsequently mapped onto the Floquet lattice. To extract the Chern number of our Floquet-version model, we extend both the energy pumping method~\cite{Martin_Phys.Rev.X_2017} and the topological oscillation method~\cite{Boyers_Phys.Rev.Lett._2020} to accommodate the 4-level quantum system. Furthermore, numerical simulations are performed to validate both methods for our 4-band Floquet half-BHZ model. The simulation results demonstrate the successful extraction of the Chern number using either of these methods, exhibiting excellent agreement with the theoretical phase diagram provided by the original model. Moreover, by employing the extended energy pumping techniques, we also extract 2D phase diagrams for our 4-band model. Additionally, we briefly discuss a potential experimental scheme for implementing our model on quantum computing platforms.

\begin{figure}[h]
\includegraphics{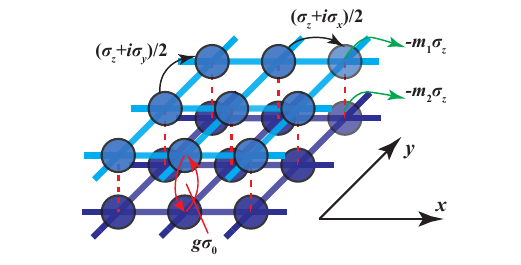}
\caption{\label{fig:epsart}The real-space lattice of the bilayer half-BHZ Model. The hopping motions along two positive directions within the upper layer are depicted by black arrows, each annotated with its corresponding strength, while the hopping behavior inside the lower layer mirrors that of the upper layer; the onsite potentials of both layers are connected by green lines, where distinct effective mass terms $m_1\sigma_z$ and $m_2\sigma_z$ are applied to each layer; the interlayer hopping is represented by the red lines, with a strength of $g\sigma_0$ in both directions.}
\end{figure}

\vspace{5pt}
{\it Bilayer Half-BHZ} 4{\it-band Model.} The half-BHZ model, also referred to as the Qi-Wu-Zhang (QWZ) model, is a two-band tight-binding Hamiltonian employed for characterizing a Chern insulator with a Chern number of $C=\pm1$. To construct a model with higher Chern numbers, the dimension of internal freedom or the band number should be increased. In this study, we propose a model by coupling two half-BHZ lattices as illustrated in Fig.~1, which provides an augmented degree of internal freedom compared to the original one. The real-space tight-binding Hamiltonian for our proposed model is given by
\begin{equation}\begin{aligned}
&\mathcal{H}_{\rm real} \\ 
&= \sum_{i,j} \Big[c_{i+1,j}^\dagger(\frac{\sigma_z+i\sigma_x}{2}\oplus\frac{\sigma_z+i\sigma_x}{2})c_{i,j}+{\rm h.c.}\Big] \\ 
&+\sum_{i,j} \Big[c_{i,j+1}^\dagger(\frac{\sigma_z+i\sigma_y}{2}\oplus\frac{\sigma_z+i\sigma_y}{2})c_{i,j}+{\rm h.c.}\Big] \\ 
&-\sum_{i,j}\left\{ c_{i,j}^\dagger \big[(m_1\sigma_z \oplus m_2 \sigma_z)-g(\sigma_x\otimes \sigma_0)\big]c_{ij}  \right\},
\end{aligned}\end{equation}    
where $\sigma_x$, $\sigma_y$ and $\sigma_z$ are the Pauli operators, $\sigma_0$ is the identity operator; the $c^\dagger_{i,j}$ ($c_{i,j}$) represents the fermion creation (annihilation) operator at the $(i,j)$-th lattice site; the first and second terms of this equation represent hopping in the $x$ and $y$ directions, respectively, while $\rm{h.c.}$ denotes Hermitian conjugation; the last term denotes the onsite potentials for both layers (the term $m_1\sigma_z\oplus m_2\sigma_z$) and the inter-layer hopping (the term $g(\sigma_x\otimes \sigma_0)$); the symbols $m_1$ and $m_2$ respectively denote the effective mass of the fermions in the two layers, while $g$ is defined as the inter-layer hopping strength (the intra-layer hopping strength is rescaled to 1 for simplicity).

Choosing $\left\{|1,A\rangle,|1,B\rangle,|2, A\rangle,|2,B\rangle\right\}$ as a complete basis for the internal freedom, where $|n,X\rangle$ ($n\in \left\{1,2\right\}$,$X\in\left\{A,B\right\}$) stands for the state located in orbit $X$ of the $n$-th layer, then the hopping inside the internal subspace could be represented as $4\times 4$ matrix. Subsequently, we perform the Fourier transformations to the fermion field operators:
\begin{equation}\begin{aligned}
c^\dagger_{i,j} &=\sum_{\mathbf{k}}e^{-i\mathbf{k}\cdot(i,j)}c^\dagger_{\mathbf{k}} \\
c_{i,j} &=\sum_{\mathbf{k}}e^{i\mathbf{k}\cdot(i,j)}c_{\mathbf{k}},
\end{aligned}\end{equation}
where $\mathbf{k}=(k_x,k_y)$ is the wavevector, and the lattice constant $a=1$. We then obtain the Hamiltonian in the momentum space $\mathcal{H}_{\mathbf{k}}= c_{\mathbf{k},\alpha}^\dagger h^{\alpha\beta}_{\mathbf{k}} c_{\mathbf{k},\beta}$, and the matrix representation of coefficients $h^{\alpha \beta}_{\mathbf{k}}$ reads
\begin{widetext}
\begin{equation}
h(\mathbf{k}) = \begin{bmatrix}-m_1+\cos k_x+\cos k_y&\sin k_x-i\sin k_y&g&0\\ \sin k_x+i\sin k_y&m_1-\cos k_x-\cos k_y&0&g\\ g&0&-m_2+\cos k_x+\cos k_y&\sin k_x- i\sin k_y\\ 0&g& \sin k_x+i\sin k_y& m_2-\cos k_x-\cos k_y\end{bmatrix}.
\label{eq:wideeq}
\end{equation}
\end{widetext}

By utilizing the momentum-space Hamiltonian, we can determine the insulating phases in our model by evaluating Chern numbers defined as $C=(1/2\pi)\sum_{i=1,2}\int_{\rm BZ}\mathcal{B}_{i} dk_x\wedge dk_y$, where $\rm BZ$ represents the first Brillouin zone, $\mathcal{B}_{i}$ denotes the $z$ component of Berry curvature for the $i$-th band, and $\sum_{i=1,2}$ signifies summation over occupied bands. Moreover, semimetallic phases are present at the boundaries between distinct insulating phases.

\begin{figure}[h]
\includegraphics{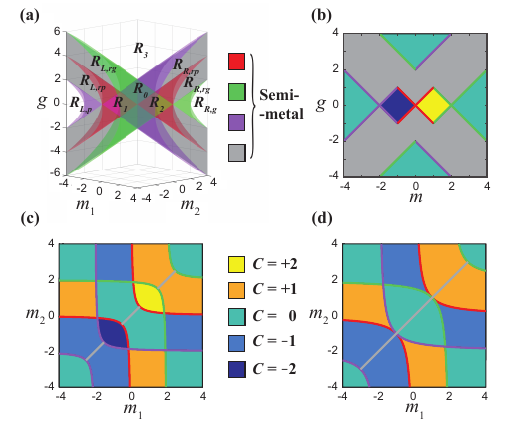}
\caption{\label{fig:epsart}Phase diagrams of the bilayer half-BHZ Model. (a) The phase boundaries in the parameter space, where the red, green, and purple cone surfaces correspond to $m_1m_2=g^2$, $(m_1-2)(m_2-2)=g^2$, and $(m_1+2)(m_2+2)=g^2$ respectively, the ``plane'' $m_1=m_2$ is painted grey, the regions cut by them are labeled. (b) 2D phase diagram when $m\equiv m_1=m_2$. (c) 2D phase diagram when $g=0.5$. (d) 2D phase diagram when $g=1.0$. (Distinct colors are employed to label different phases or phase boundaries in all subfigures, as illustrated on the right side of (a) and (b).)}
\end{figure}

As shown in Fig.~2(a), the boundaries within the phase diagram can be defined by three cone surfaces and a plane in the parameter space $(m_1,m_2,g)$. The boundaries are characterized by the following equations: $m_1m_2=g^2$, $(m_1-2)(m_2-2)=g^2$, $(m_1+2)(m_2+2)=g^2$, and $m_1=m_2$, respectively. The three cone surfaces correspond to the Dirac semimetal phase, wherein Dirac cones emerge at high-symmetry points in the Brillouin zone [i.e., $(\pi,0)$, $(0,0)$, and $(\pi,\pi)$ for the three cone surfaces, respectively], representing well-established topological phase transitions. The plane defined by $m_1=m_2$ also represents a semimetallic phase; however, in this case, the conduction band touch the valence band along lines rather than isolated points. Additionally, the energy gap closes in this situation, indicating a phase transition as well. Importantly, it should be noted that the semimetallic phase within the $m_1=m_2$ plane exists only within a specific region depicted in Fig.~2(b).
 
For the remaining insulating phases delineated by the aforementioned boundaries, we assign Chern numbers to classify them. For convenience, we define the regions in Fig.~2(a) as follows: Firstly, we define the spindle enclosed by the right half of the purple cone and the left half of the green cone as $R_0$. Within $R_0$, two smaller spindles are determined by the red cone; we label the left one as $R_1$ and the right one as $R_2$. In addition to these middle regions, we designate the region between the left half of the green cone and the left half of red cones as $R_{\rm L,rg}$ and its counterpart on right side as $R_{\rm R,rg}$. Similarly, regions $R_{\rm L,rp}$, $R_{\rm L,p}$, $R_{\rm R,rp}$ and $R_{\rm R,g}$ are defined accordingly. Finally, the only remaining region outside all cones is defined as $R_3$. With well-defined regions established, their corresponding Chern numbers are given in Tab. 1. To provide a direct visualization of our phase diagram, we also present 2D phase diagrams with $m\equiv m_1=m_2$, $g=0.5$, and $g=1$ in Figs.~2(b), (c) and (d) correspondly.

\begin{table}[b]
\caption{\label{tab:Tab1}
Chern numbers of different regions in the parameter space (all semimetal boundaries are not included).}
\begin{ruledtabular}
\begin{tabular}{cccc}
Region & \multicolumn{1}{c}{ Chern number} & Region & Chern number\\
\hline
$R_0\backslash(R_1\cup R_2)$ & \multicolumn{1}{c}{0} & $R_{\rm L,rp}\backslash R_0$ & $-1$\\
$R_1$ & \multicolumn{1}{c}{$-2$} & $R_{\rm L,rg}\backslash R_0$ & $+1$\\
$R_2$ & \multicolumn{1}{c}{$+2$} & $R_{\rm R,rp}\backslash R_0$ & $-1$ \\
$R_3$ & \multicolumn{1}{c}{0} & $R_{\rm R,rg}\backslash R_0$ & $+1$ \\
$R_{\rm L,p}$ & \multicolumn{1}{c}{0} & $R_{\rm R,g}\backslash R_0$ & $0$ \\
\end{tabular}
\end{ruledtabular}
\end{table}

\vspace{5pt}
{\it Floquet-version} 4{\it-band Model.} To simulate the bilayer half-BHZ model proposed above, it is crucial to map this model onto the Synthetic Floquet lattice, and we refer to the mapped model as the Floquet-version 4-band model. In order to elucidate the mapping process, we introduce the concept of the Floquet lattice and the Floquet quantum system first.

A Floquet quantum system with $N$ drives can be described by the time-dependent Hamiltonian $\mathcal{H}(\vec{\omega}t+\vec{\phi})$, where $\vec{\omega}\equiv(\omega_1,\omega_2,\cdots,\omega_N)$ and $\vec{\phi}\equiv(\phi_1,\phi_2,\cdots,\phi_N)$ are vectors containing the frequencies and initial phases of the $N$ drives. By choosing $\Gamma\equiv\left\{|\gamma\rangle\right\}$ as a complete basis for this quantum system, the Hamiltonian (with the Einstein summation convention) expands into
\begin{equation}
\mathcal{H}^{\alpha \beta} (\vec{\omega}t+\vec{\phi})\vert \alpha\rangle\langle \beta|, \text{ with } \vert \alpha\rangle\text{, }\vert \beta\rangle \in \Gamma.
\end{equation}
According to the Floquet-Bloch theory~\cite{Gomez_PRL_2013}, the eigenstate of the above Hamiltonian is given by
\begin{equation}
|\psi(t)\rangle = e^{-i\epsilon t/\hbar}\varphi^\alpha(t)|\alpha\rangle,
\end{equation}
where $\epsilon$ represents the Floquet quasi-energy. We expand the coefficients $\mathcal{H}^{\alpha \beta}(\vec{\omega}t+\vec{\phi})$ and $\varphi^\alpha(t)$ into Fourier series $\sum_{\vec{n}}h^{\alpha\beta}_{\vec{n}}e^{-i\vec{n}\cdot(\vec{\omega}t+\vec{\phi})}$, $\sum_{\vec{n}}\varphi^\alpha_{\vec{n}}e^{-i\vec{n}\cdot\vec{\omega}t}$, and substitute into the Schr\"{o}dinger equation 
\begin{equation}
i\hbar \partial_t |\psi(t)\rangle= \mathcal{H}(\vec{\omega}t+\vec{\phi})|\psi(t)\rangle,
\end{equation}
then the eigenequation can be modefied to
\begin{widetext}
\begin{equation}
\sum_{\vec{m}}( h_{\vec{m}}^{\alpha\beta}e^{-i\vec{m}\cdot(\vec{\omega} t+\vec{\phi})}|\alpha\rangle\langle\beta|)\sum_{\vec{n}} (e^{-i(\vec{n}-\vec{m})\cdot \vec{\omega}t}\varphi^\gamma_{\vec{n}-\vec{m}}|\gamma\rangle) -\sum_{\vec{n}}\vec{n}\cdot\hbar\vec{\omega} e^{-i\vec{n}\cdot\vec{\omega}t}\varphi^\gamma_{\vec{n}}|\gamma\rangle =\epsilon \sum_{\vec{n}}e^{-i\vec{n}\cdot\vec{\omega}t}\varphi^\gamma_{\vec{n}}|\gamma\rangle,
\label{eq:wideeq}
\end{equation}
\end{widetext}
where the phase factor $e^{im_i\omega_it}$ ($e^{-im_i\omega_it}$) represents the process of emitting (absorbing) $m_i$ ($m_i>0$) photons to (from) the drive with frequency $\omega_i$. In the so-called ``Floquet representation"~\cite{Rudner_arXiv_2020}, we define the Fock state $|\vec{m}\rangle\equiv |m_1,m_2,\cdots,m_N\rangle$ as the $\vec{m}$-th harmonic of Fourier expansion of $|\psi(t)\rangle$. The phase factors can be expressed as 
\begin{equation}\begin{aligned}
e^{im_i\omega_it} & \equiv(a_{i})^{m_i}= |n_i-m_i\rangle\langle n_i| \\
e^{-i{m_i}\omega_it} & \equiv(a_{i}^\dagger)^{m_i}= |n_i+m_i\rangle\langle n_i|
\end{aligned}\end{equation}
Based on the Floquet representation, we can further derive an effective Hamiltonian $\mathcal{H}_{\rm eff}$ from Eq.~(7), which reads
\begin{equation}\begin{aligned}
\mathcal{H}_{\rm eff} &=\sum_{\vec{n},\vec{m}} h^{\alpha\beta}_{\vec{m}}e^{-i\vec{m}\cdot\vec{\phi}} (|\vec{n}\rangle\langle \vec{n}-\vec{m}|\otimes|\alpha\rangle\langle \beta|) \\&-\sum_{\vec{n}}\vec{n}\cdot \hbar\vec{\omega} |\vec{n}\rangle\langle \vec{n}|.
\end{aligned}\end{equation}
If we consider $|\vec{n}\rangle$ as a lattice site, the set $\left\{|\vec{n}\rangle\right\}$ encompasses all possible sites and forms the Floquet lattice with dimension $N$. In the case of incommensurate drives, the Floquet lattice becomes infinite, and $\mathcal{H}_{\rm eff}$ is equivalent to a tight-binding Hamiltonian defined on the Floquet lattice. 

In the tight-binding Hamiltonian, the phase factor $e^{-i\vec{m}\cdot\vec{\phi}}$ in $\mathcal{H}_{\rm eff}$ induces an effective vector potential $\vec{\phi}$ according to Peierls substitution. The second term, $-\vec{n}\cdot \hbar \vec{\omega}$ represents an effective electric potential acting on the lattice with $\hbar\vec{\omega}$ denoting the uniform field strength. Moreover, It should be noted that the Floquet lattice is only valid in the strong driving limit~\cite{Martin_Phys.Rev.X_2017, Crowley_PRB_2019} (or equivalently, in the adiabatic limit). Specifically, for any two instantaneous eigenenergies, their minimal difference denoted as $\Delta$ should be much larger than the energy scale of all drives $\hbar \omega_i$ ($i=1,2,\cdots,N$).  

Based on the Floquet lattice introduced above, we derive obtain the Floquet-version time-dependent Hamiltonian $\mathcal{H}_{\rm F}(t)$ by substituting the wavevector $\mathbf{k}$ in the original momentum-space Hamiltonian with $\vec{\omega} t+\vec{\phi} =(\omega_1 t+\phi_1,\omega_2 t+\phi_2)$. Under the complete basis set $\left\{1A, 1B, 2A, 2B \right\}$, $\mathcal{H}_{\rm F}(\vec{\omega}t+\vec{\phi})$ transforms into a $4\times 4$ matrix, which reads
\begin{equation}
\mathcal{H}_{\rm F}(\vec{\omega}t+\vec{\phi})= \eta h({\vec{\omega}t+\vec{\phi}}),
\end{equation}
where $h$ has been defined in Eq.~(3). Besides, It is noteworthy that we also introduce a dimensionless coefficient, denoted as $\eta$, in order to ensure compliance with the strong driving limit.

Furthermore, we transform the time-dependent Hamiltonian $\mathcal{H}_{\rm F}(\vec{\omega}t+\vec{\phi})$ into its Floquet representation using Eq.(8). The resulting effective Hamiltonian on the Floquet lattice, denoted as $\mathcal{H}_{\rm FL}$, which reads
\begin{equation}\begin{aligned}
&\mathcal{H}_{\rm FL}\\
&=\eta\left(-\frac{m_1+m_2}{2}\sigma_0\otimes\sigma_z+\frac{m_2-m_1}{2}\sigma_z\otimes\sigma_z\right) \\ 
&+\eta\left[\frac12e^{i\phi_1}(\sigma_0\otimes \sigma_z -i\sigma_0\otimes\sigma_x)a_1+{\rm h.c.}\right] \\ 
&+\eta\left[\frac12e^{i\phi_2}(\sigma_0\otimes \sigma_z -i\sigma_0\otimes\sigma_y)a_2+{\rm h.c.}\right] \\ 
&+\eta\left[g(\sigma_x\otimes \sigma_0)\right]-\vec{n}\cdot\hbar\vec{\omega}.
\end{aligned}\end{equation}
The Hamiltonian $\mathcal{H}_{\rm FL}$ is mathematically equivalent to $\mathcal{H}_{\rm real}$ and shares the same phase diagram as described in the previous section. Moreover, as a specific example of $\mathcal{H}_{\rm eff}$, $\mathcal{H}_{\rm FL}$ is augmented by a $1+2$ dimensional vector potential $A_{\tau}=(-\vec{n}\cdot\hbar\vec{\omega}, \vec{\phi})$, where the index $\tau$ represents the $\tau$-th component of the $1+2$-dimensional vector ($\tau=0,1,2$; with ``0'' denoting the time dimension, ``1'' and ``2'' corresponding to the synthetic spatial dimension induced by drives with frequencies $\omega_1$ and $\omega_2$, respectively). The vector potential $\vec{\phi}$ remains constant throughout the lattice in $\mathcal{H}_{\rm FL}$ and does not contribute to the effective electromagnetic field, while the scalar potential $-\vec{n}\cdot \hbar\vec{\omega}$ generates a uniform electric field.

\vspace{5pt}
{\it Quantized Energy Pumping.} In this section, we extend the (quantized) energy pumping approach proposed by Martin \textit{et al.} in Ref.~\onlinecite{Martin_Phys.Rev.X_2017} to extract the Chern number of the Floquet-version 4-band model derived in the preceding section and present our numerical simulation results.

As previously discussed, our Floquet-version 4-band model exhibits an effective vector potential $A_\tau=(-\vec{n}\cdot \hbar \vec{\omega},\vec{\phi})$, analogous to the case of the integer quantum Hall effect in real space. The motion of a fermion on the Floquet lattice under the influence of $A_\tau$ is governed by a semiclassical equation~\cite{Sundaram_PRB_1999}. Furthermore, based on this semiclassical equation, Martin \textit{et al.} propose the energy pumping approach in Ref.~\onlinecite{Martin_Phys.Rev.X_2017}.

In contrast to the derivation presented in Ref.~\onlinecite{Martin_Phys.Rev.X_2017}, we provide a simplified derivation of the energy pumping effect for our 4-band model utilizing the $1+2$ dimensional Chern-Simons insulator theory. In this theory, the $1+2$ dimension current vector reads~\cite{Qi_RMP_2011}:
\begin{equation}
j_\mu^i = \frac{C_i}{2\pi}\epsilon^{\mu\nu\tau}\partial_\nu A_\tau,
\end{equation}
where the indice $i$ denotes the $i$-th band ($i=1,2,3,4$, and the four bands are arranged in descending order according to the magnitude of their eigenenergies), $C_i$ is the Chern number of the $i$-th band, $\epsilon^{\mu\nu\tau}$ is the Levi-Civita symbol, with $\mu$, $\nu$, and $\tau$ indices representing the component of $1+2$ vector or tensor ($\mu,\nu,\tau=0,1,2$). Again, the Einstein summation convention is employed. 

By substituting $A_\tau$ into Eq.~(12), we obtain $j_{\mu}^i = C_i(0, \hbar\omega_2,\hbar\omega_1)/2\pi$. The total current vector $j_\mu$ is the summation of the occupied bands, which reads
\begin{equation}
j_\mu = \sum_{i=1,2} j^i_\mu = \frac{C}{2\pi}(0,-\hbar\omega_2,\hbar\omega_1), 
\end{equation}
where $C= C_1+C_2$ denotes the total Chern number. 

According to the definition of the Floquet representation, the hopping process along the positive direction of the $m$-th synthetic spatial dimension in the Floquet lattice is denoted as $a^\dagger_m$ ($m=1,2$), which corresponds to the absorption of a photon with energy $\hbar \omega_{m}$. Consequently, the energy flux along the $m$-th dimension of the Floquet lattice can be expressed as $\hbar\omega_m j_m$, wherein $j_m$ denotes the synthetic spatial component of $j_\mu$. By defining the energy associated with the $m$-th drive as $E_m\equiv \int_0^t \hbar\omega_m j_m(t^\prime) dt^\prime$, we obtain the pumping rates (the energy fluxes), which reads
\begin{equation}\begin{aligned}
\frac{dE_1}{dt}&=-\frac{C\hbar^2\omega_1\omega_2}{2\pi}\\
\frac{dE_2}{dt}&=\frac{C\hbar^2\omega_1\omega_2}{2\pi}.
\end{aligned}\end{equation}
This equation illustrates the energy conversion between two drives, where the discrete Chern number $C$ quantizes the pumping rates. Therefore, we could extract the Chern number $C$ through 
\begin{equation}
C=\sum_{n=1}^2\pi\left(\frac{dE_2^n}{dt}-\frac{dE_1^n}{dt}\right)/(\hbar^2\omega_1\omega_2),
\end{equation}
The equation presented here is a straightforward derivation of Eq.~(14), where we decompose the energy pumping rates ($dE_m/dt$, with $m=1,2$ denoting the respective drives) into the summation of occupied bands ($\sum_{n=1,2}dE^n_m/dt$, with $n=1,2$ representing the corresponding bands) to facilitate subsequent numerical simulations.  Furthermore, it is important to note that this quantized pumping phenomenon is meaningful only when the energies are averaged over a large time scale. This requirement can be attributed to an additional term in the semiclassical motion equation mentioned in Refs.~\onlinecite{Martin_Phys.Rev.X_2017,Sundaram_PRB_1999}.

To validate the energy pumping approach described above, we conduct numerical simulations on our Floquet-version time-dependent Hamiltonian $\mathcal{H}_{\rm F}(\vec{\omega}t+\vec{\phi})$ (Eq.~(7)). This includes three processes. Firstly, we construct the initial state $|\psi^n(0)\rangle=|u_n(\vec{\phi})\rangle$, where $|u_n(\vec{\phi})\rangle$ is the $n$-th eigenstate of $\mathcal{H}_{\rm F}(\vec{\phi})$ ($n=1,2$). Secondly, we evaluate the evolution operator by multiplying a series of successive ``infinitesimal" evolution operators $\exp(-i\mathcal{H}_{\rm F}(\vec{\omega}t+\vec{\phi})\delta t)$, where $\delta t$ denotes the infinitesimal time interval and we adopt $\hbar=1$ throughout the simulation. The expression of the evolution operator from $t^\prime=0$ to $t^\prime=N\delta t$ is given by 
\begin{equation}
\mathcal{U}(N\delta t) \approx \prod_{j=0}^{N-1} \exp\left[-i\mathcal{H}_{\rm F}(\vec{\omega} j \delta t+\vec{\phi})\delta t\right],
\end{equation}
and the accuracy of this evaluation could be ensured by taking $\delta t=10^{-3}$. Finally, the discretized evolution trajectory of the quantum state $\left\{|\psi^n(j\delta t)\rangle\right\}$ is obtained by applying the unitary operator $\mathcal{U}(j\delta t)$ ($j=0,1,\cdot,N$) to $|\psi^n(0)\rangle$. Moreover, to satisfy the strong driving limit (i.e., $\eta\Delta \gg \omega_1,\omega_2$) and meet the requirement for a large time scale, we set the total simulation time $T\equiv N\delta t=10^5$, where frequencies $\omega_1=0.1$ and $\omega_2 =\gamma\omega_1$ ($\gamma=(\sqrt{5}-1)/2$ represents the golden ratio) are chosen.

After the numerical simulation,  post-data processing is required to extract the pumping rate $dE_m^n/dt$ ($m,n=1,2$) in Eq.~(15) for subsequent calculation of the Chern number. We perform the post-data processing in five steps. Firstly, we partition $\mathcal{H}_{\rm F}(\vec{\omega}t +\vec{\phi})$ into two components, $\mathcal{H}_1(\omega_1,\phi_1,t)$ and $\mathcal{H}_2(\omega_2,\phi_2,t)$, based on the drive frequencies while disregarding the time-independent portion. The expressions for these components are as follows 
\begin{equation}\begin{aligned}
\mathcal{H}_1(\omega_1,\phi_1,t) &=  \sin(\omega_1 t+\phi_1)\sigma_0\otimes\sigma_x \\ &+ \cos(\omega_1 t+\phi_1) \sigma_0\otimes\sigma_z\\
\mathcal{H}_2(\omega_2,\phi_2,t) &=  \sin(\omega_2 t+\phi_2)\sigma_0\otimes\sigma_y \\ &+ \cos(\omega_2 t+\phi_2) \sigma_0\otimes\sigma_z.
\end{aligned}\end{equation}
Secondly, we calculate the instant energy pumping rates as $dE_m^n(j\delta t)/dt = \langle \dot{\mathcal{H}}_m(j\delta t)\rangle_{j,n}$. Here, $\dot{\mathcal{H}}_m$ denotes the time derivative of $\mathcal{H}_m$, and $\langle O\rangle_{j,n} \equiv \langle \psi^n(j\delta t) |O|\psi^n(j\delta t)\rangle$. Thirdly, we perform numerical integration using $\left\{dE_m^n(j\delta t)/dt\right\}$ to obtain the enregy $\left\{E_m^n(j\delta t)\right\}$. Fourthly, we perform linear regression on the set $\left\{E_m^n(j\delta t)\right\}$ using linear functions $y=a_{m,n}x+b_{m,n}$ respectively. Finally, we interpret the slope $a_{m,n}$ of the regression function as the rates of energy transfer $dE_m^{n}/dt$, and subsequently extract the Chern number using Eq.~(15).

\begin{figure}[h]
\includegraphics{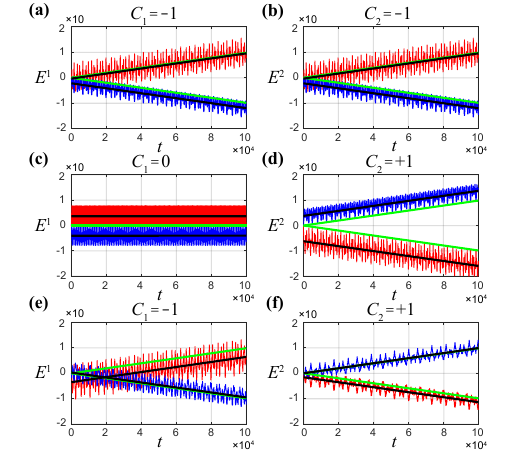}
\caption{\label{fig:epsart} The Simulation results of the energy pumping. The red and blue lines correspond to the energies of the first and second drives (i.e., $E^n_m$), respectively, while the black lines represent fitted energy pumping curves. The green lines depict the theoretical predictions of energy pumping. The superscript $n$ in $E^n$ represents the $n$-th band ($n=1,2$). In (a) and (b), $(m_1,m_2,g)=(-1,-1,0.5)$, $C=-2$. In (c) and (d), $(m_1,m_2,g)=(1,-6,0.5)$, $C=1$. In (e) and (f), $(m_1,m_2,g)=(0.5,-1.4,0.5)$, $C=0$. (The value of $\eta=4$ is chosen to ensure the fulfillment of the strong driving limit.)}
\end{figure}

The simulation results for quantized energy pumping are presented in Fig.~3. In Figs.~3(a) and (b), we set parameters $(m_1,m_2,g)=(-1,-1,0.5)$, which correspond to a topological phase with $C=-2$. The fitted energy pumping lines in Fig.~3(a) yield a Chern number of approximately $-1$ for the first band (i.e., $C_1\approx -1$), while fitted lines in Fig.~3(b) give a Chern number of approximately $-1$ for the second band (i.e., $C_2\approx -1$); both agree well with the phase diagram described in the first section and provide a reliable estimate of the total Chern number $C$. Furthermore, to validate the universality of our energy pumping methods, we also perform simulations on a topological phase with $C=+1$ in Figs~.3(c) and (d), as well as a trivial phase ($C=0$) with $C_1=-C_2=-1$, depicted in Figs.~3(e) and (f). The simulation results are all consistent with theoretical predictions from our phase diagram. It is noteworthy that while the slopes of all fitted curves accurately align with their respective theoretical predictions, certain fitted curves exhibit non-zero intercepts due to the limited accuracy of numerical integration near $t^\prime =0$. Nevertheless, this discrepancy does not have a significant impact on our estimation.

\begin{figure}[h]
\includegraphics{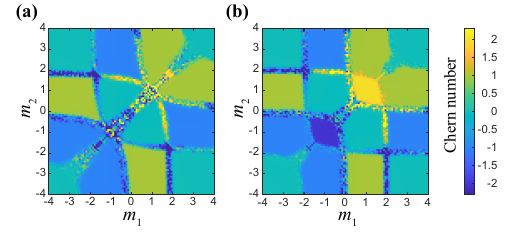}
\caption{\label{fig:epsart} The simulated 2D phase diagrams. (a) Simulated phase diagram with $g=1$ whose theoretical counterpart is shown in Fig.~2(d). (b) Simulated phase diagram with $g=0.5$, its theoretical counterpart is given in Fig.~2(c). (The color bar on the right side of (b) represents the simulated Chern number, which is applicable to both subfigures. Additionally, we set $\eta=4$ during the simulation.)}
\end{figure}

Moreover, employing the approach of quantized energy pumping, we have also constructed two simulated 2D phase diagrams (Fig.~4) by uniformly sampling $101\times 101$ points within the region defined by $-4\leq m_1\leq 4$ and $-4\leq m_2\leq 4$, while setting $g=1$ for Fig.~4(a) and $g=0.5$ for Fig.~4(b). Both diagrams exhibit remarkable consistency with their corresponding theoretical counterparts (The theoretical predictions are presented in Fig.~2(d) and Fig.~2(c), respectively.), except for slight deviations observed near the phase boundaries due to violation of strong driving limit imposed by a narrow energy gap.

\vspace{5pt}
{\it Topological Oscillation.} In the energy pumping approach, we extract the Chern number from the ``real space'' of the Floquet lattice. Additionally, another method known as ``topological oscillation'' can also derive the Chern number from the ``momentum space'' of the Floquet lattice~\cite{Boyers_Phys.Rev.Lett._2020}. In this section, we extend this approach to extract the Chern number of our Floquet-version 4-band model as a complement to the energy pumping method.

As demonstrated in the second section, the wavevector $\mathbf{k}$ in the Bilayer half-BHZ model is substituted by $\vec{\omega}t+\vec{\phi}$ to derive the Floquet-version 4-band model. Consequently, it becomes evident that the momentum space of the Floquet lattice corresponds to the ``frequency space'', with $\vec{\Omega}(t)\equiv\vec{\omega} t+\vec{\phi}$ playing an equivalent role as $\mathbf{k}$. Analogous to the Brillouin zone (BZ), we can define a Floquet Brillouin zone~\cite{Rudner_arXiv_2020} (FBZ) for our model's frequency space, which reads
\begin{equation}
{\rm FBZ}=\left\{\vec{\Omega}|-\pi<\Omega_i\leq\pi,i=1,2 \right\}.
\end{equation}

After defining the FBZ, we return to the topic of topological oscillations. It is noteworthy that in the strong driving limit (i.e., $\eta\Delta \gg \hbar\omega_1,\hbar\omega_2$) is equivalent to the adiabatic limit of quantum dynamics, as the large energy gap $\eta\Delta$ ensures negligible excitation probability between different instantaneous eigenstates and each eigenstate evolves nearly adiabatically. To illustrate, assuming the intial quantum state as $|\psi(0)\rangle=\sum_{j=1}^4 a_j |u_j(\vec{\phi})\rangle$, where $|u_j(\vec{\phi})\rangle$ represents the $j$-th eigenstate of $\mathcal{H}_{\rm F}(\vec{\phi})$ and $\sum_{j=1}^4 |a_j|^2=1$, thus, in the adiabatic limit, the state evolves to
\begin{equation}\begin{aligned}
&|\psi(t)\rangle=\\ & \sum_{j=1}^4 e^{-\frac{i}{\hbar}\int_0^t E_j(t^\prime)dt^\prime +i\int_l \vec{\mathcal{A}}_j\cdot d\vec{l}} |u_j(\vec{\Omega}(t))\rangle,
\end{aligned}\end{equation}
after time $t$, where $\vec{\mathcal{A}}_j$ represents gauge-dependent Berry connection defined as $i\langle u_j(\vec{\Omega})|\nabla_{\vec{\Omega}}|u_j(\vec{\Omega})\rangle$ with $\vec{\Omega}\in {\rm FBZ}$; $l$ represents the trajectory of $\vec{\Omega}(t)$ in FBZ from $t^\prime =0$ to $t^\prime =t$.

In Eq.~(19), the adiabatic drift of $\vec{\Omega}(t)$ is induced by the unified electric field $\vec{\omega}$, and with sufficiently long evolution time, its trajectory $l$ can span across the entire FBZ due to the incommensurate nature between different drives. It is noteworthy that apart from the normal dynamical phase $\exp(-i\int_0^t E_j(t^\prime)dt^\prime/\hbar)$, the geometrical phase $\exp(i\int_l\vec{\mathcal{A}}_j\cdot d\vec{l})$ is also included. Therefore, it becomes possible to extract the Chern number through a carefully designed quantum dynamics process (the Chern number could also be defined as $C=(1/2\pi)\sum_{j=1,2}\int_{\rm \partial FBZ} \vec{\mathcal{A}}_j\cdot d\vec{l}$, where $\partial {\rm FBZ}$ denotes the boundary of FBZ.).
 
The above assumption could be achieved by measuring the oscillation of overlap $F(t)=|\langle \psi(t)|\psi^\prime(t)\rangle|^2$~\cite{Boyers_Phys.Rev.Lett._2020}, where $|\psi(t)\rangle=\mathcal{U}(t)\sum_{j=1}^N a_j |u_j(\vec{\phi})\rangle\equiv\mathcal{U}(t)|\psi(0)\rangle$, $|\psi^\prime(t)\rangle=\mathcal{U}(t)\sum_{j=1}^N a_j |u_j(\vec{\phi}+\vec{\delta\phi})\rangle\equiv\mathcal{U}(t)|\psi^\prime(0)\rangle$, $\mathcal{U}(t)$ is the evolution operator from $t^\prime =0$ to $t^\prime =t$, $\vec{\delta\phi}$ is a small quantity satisfying $\vec{\omega}\nparallel \vec{\delta \phi}$. Take $N=4$, we extended the method to our 4-band model, and the overlap $F(t)$ reads
\begin{equation}
F(t) \simeq 1-4\sum_{1\leq i<j\leq 4}\sin^2\left[(C_i-C_j)\omega_{\rm T} t/2\right]P_iP_j,
\end{equation}
where $\omega_{\rm T}\equiv |\vec{\delta\phi}\times \vec{\omega}|/2\pi$, $P_i\equiv |\langle u_i(\vec{\phi})|\psi(0)\rangle|^2$, and this equation holds approximately when neglecting the high-order terms of $\vec{\delta\phi}$. However, the Chern number $C=C_1+C_2$ is absent in Eq.~(20), which makes the original topological oscillation method in Ref.~\onlinecite{Boyers_Phys.Rev.Lett._2020} ineffective. To address this limitation, we propose a two-step approach.

In the first step of the two-step approach, we prepare the initial states $|\psi(0)\rangle$ and $|\psi^\prime(0)\rangle$ as equally weighted superpositions of four eigenstates ($\sum_{j=1}^4 |u_j(\vec{\phi})\rangle/\sqrt{4}$ and $\sum_{j=1}^4 |u_j(\vec{\phi}+\vec{\delta\phi})\rangle/\sqrt{4}$), followed by recording both dynamic processes. Subsequently, based on Chern numbers of all four bands ($\mathcal{C}\equiv(C_1,C_2,C_3,C_4)$), we evaluate the first overlap function $F_1(t)$, which exhibits distinct oscillation modes as follows 
\begin{equation}\begin{aligned}
&F_1(t)= \\ &\begin{cases} \cos^2(\omega_{\rm T} t) & \rm \mathcal{C}\in\rm \left\{Per(-1,-1,+1,+1)\right\} \\
\cos^4(\omega_{\rm T} t/2) & \rm \mathcal{C}\in\rm \left\{Per(-1,0,+1,0)\right\}\\
1 &\rm \mathcal{C}=(0,0,0,0)
\end{cases},\end{aligned}
\end{equation}
where the $\rm \left\{Per(\cdot,\cdot,\cdot,\cdot)\right\}$ represents the set comprising all possible permutations of the array $(\cdot,\cdot,\cdot,\cdot)$. From Eq.~(21), it is evident that $F_1=\cos^4(\omega_{\rm T} t/2)$ implies $C=\pm 1$, and $F_1(t)=1$ indicates $C=0$. However, $F_1(t)=\cos^2(\omega_{\rm T} t)$ does not indicate $C=\pm2$, since some permutations of $(-1,-1,+1,+1)$ like $(-1,+1,+1,-1)$ indicate topological trivial phase ($C=0$). Therefore, an additional step is necessary when $F_1(t)=\cos^2(\omega_{\rm T} t)$.

In the second step, we prepare the initial states $|\psi(0)\rangle$ and $|\psi^\prime(0)\rangle$ as equally weighted superpositions of the two eigenstates with lower eigenenergies ($\sum_{j=1,2}|u_j(\vec{\phi})\rangle/\sqrt{2}$ and $\sum_{j=1,2}|u_j(\vec{\phi}+\vec{\delta\phi}\rangle/\sqrt{2}$) respectively. Following the quantum dynamics process, we can evaluate the second overlap $F_2(t)$, which reads
\begin{equation}
F_2(t)=\begin{cases} 1 & C=\pm2 \\ \cos^2(\omega_{\rm T} t) & C=0
\end{cases}.
\end{equation}
Based on Eq.~(22), the remaining issue of $F_1(t)$ can be addressed. Overall, our two-step approach enables the distinction of phases with Chern numbers $C=0$, $C=\pm 1$, and $C=\pm 2$. However, determining the sign is not available due to the even function nature of $F_i(t)$ ($i=1,2$).

\begin{figure}[h]
\includegraphics{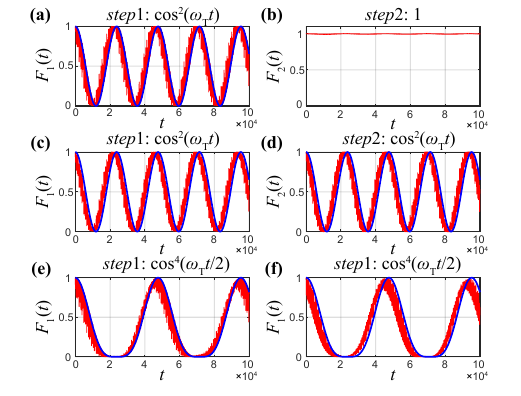}
\caption{\label{fig:epsart} The simulation results and theoretical predictions of the overlap functions ($F_1(t)$,$F_2(t)$). The simulation results are represented by the red lines, while the theoretical predictions are indicated by the blue lines. (a) and (b) depict the functions $F_1(t)$ and $F_2(t)$, respectively, obtained from the same simulation with $(m_1,m_2,g)=(1,1,0.5)$ and $C=+2$. (c) and (d) illustrate $F_1(t)$ and $F_2(t)$ of another simulation with $(m_1,m_2,g)=(0.5,-1.4,0.5)$ and $C=0$. (e) show the simulation result of $F_1(t)$ with $(m_1,m_2,g)=(-4,-1,0.5)$, $C=-1$. (f) show the simulation result of $F_1(t)$ with $(m_1,m_2,g)=(-6,1,0.5)$, $C=+1$. ($\eta =2$)}
\end{figure}

At last, we present the numerical simulation results of the aforementioned two-step approach. The simulation details and selected parameters align with those described in the energy pumping approach with $\vec{\delta \phi}=(\gamma,-1)/50$ ($\gamma=(\sqrt{5}-1)/2$). The simulation results for four different phases are depicted in Fig.~5. In Figs.~5(a) and (b), we set the parameters $(m_1,m_2,g)=(1,1,0.5)$ such that $C=+2$. The overlap function $F_1(t)\approx\cos^2(\omega_T t)$ in Fig.~5(a) and $F_2(t)\approx 1$ in Fig.~5(b) indicate $C=\pm2$ according to our two-step approach,  which aligns with the theoretical predictions. In Figs.~5(c) and (d), we choose $(m_1,m_2,g)=(0.5,-1.4,0.5)$ resulting in $C=0$. The presence of $\cos^2(\omega_T t)$ curves in both subfigures indicates a total Chern number $C=0$ with $(C_1,C_2)\in\left\{{\rm Per}(-1,1)\right\}$, which is consistent with the theoretical expectations. Furthermore, we verify topological phases with $C=+1$ and $C=-1$ in Figs.~5(e) and (f), respectively. The oscillation curve $F_1(t)\approx\cos^4(\omega_T t/2)$ indicates $C=\pm1$, which also agrees well with our two-step approach and the theoretical prediction.

\vspace{5pt}
{\it Possible Experimental implementation.} In this section, we propose a possible experimental implementation to simulate our bilayer half-BHZ model on a superconducting quantum computing system. Considering a quantum system comprising two transmon qubits~\cite{Krantz_APR_2019,Blais_RMP_2021} interconnected by a tunable inductive coupler and employing the two-level approximation, the resulting experimental Hamiltonian reads~\cite{Krantz_APR_2019}
\begin{equation}
\mathcal{H}_{\rm exp} = \sum_{i=1,2} \left[\frac{\hbar\omega_{q,i}}{2}\sigma_z^i + D^i(t)\sigma_x^i\right] + g^\prime\sigma_z^1\otimes\sigma_z^2,
\end{equation}
where $\omega_{q,i}$ denotes the frequency of $i$-th qubit; The driving term of the $i$-th qubit, denoted as $D^i(t)\sigma_x^i$, is achieved by applying microwaves on the XY control port. The coupling strength between two qubits is represented by parameter $g^\prime$, which can be tuned using the coupler.

Assuming the frequencies of both qubits are the same [i.e., $\omega_{q,1}=\omega_{q,2}\equiv \omega_0$], we could set the waveforms of driving microwave $D^i(t)$ to be $h^i_x(t)\cos(\omega_0 t-2 \int_0^t h_z^i(t^\prime) dt^\prime) - h^i_y(t)\sin(\omega_0 t-2\int_0^t h_z^i(t^\prime)dt^\prime)$ through the arbitrary waveform generator. We choose a frame work rotating with $\mathcal{H}_{\rm fra}(t)\equiv\sum_{i=1,2}(\hbar\omega_0/2-h^i_z(t))\sigma_z^i$, and introduce the transformed evolution operator $\mathcal{U}_{\rm rot}(t)\equiv\exp[-i(\int_0^t \mathcal{H}_{\rm fra}(t^\prime) dt^\prime)/\hbar]$. Consequently, the Hamiltonian in the rotating framework reads 
\begin{equation}
\begin{aligned}
\mathcal{H}_{\rm rot} &= \mathcal{U}_{\rm rot}^\dagger\mathcal{H}_{\rm exp} \mathcal{U}_{\rm rot} + i\hbar(\partial\mathcal{U}_{\rm rot}^\dagger/\partial t)\mathcal{U}_{\rm rot} \\ &= \sum_{i=1,2}\sum_{j=x,y,z} h^i_j(t) \sigma_j^i  + g^\prime \sigma_z^1\otimes\sigma_z^2.
\end{aligned}
\end{equation}
This Hamiltonian incorporates our Floquet-based model $\mathcal{H}_{\rm F}(\vec{\omega}t+\vec{\phi})$. Furthermore, a similar experimental technique has been experimentally validated in Ref.~\onlinecite{Malz_Phys.Rev.Lett._2021}. Consequently, we anticipate that the meticulous design of driving waveforms $D^i(t)$ ($i=1,2$) will enable the realization of above quantum simulation.

\vspace{5pt}
{\it Summary.} In this letter, we present an exemplary demonstration of simulating Chern insulators with diverse topological phases using the synthetic Floquet lattice. By extending both energy pumping and topological oscillation methods to the 4-band case, we showcase that the Chern number $C$ can be accurately determined through numerical simulations of the quantum dynamic process. Moreover, a potential experimental implementation is also presented.

In addition to the two-dimensional lattice, as discussed previously, the Floquet lattice enables the realization of tight-binding Hamiltonians defined on lattices of arbitrary dimensions. Consequently, synthetic Floquet lattices can be employed to simulate all non-interacting topological quantum states in the Altland-Zirnbauer periodic table~\cite{Altland_PRB_1997,Hasan_RMP_2010}. Based on previous studies and this work, we believe that the extraction techniques for other categories of topological invariants (such as the $\mathbb{Z}_2$ invariant) could be developed, thereby paving the way for a more universal quantum simulator applicable to non-interacting topological materials and facilitating the discovery of novel topological materials.

\vspace{5pt}
{\it Acknowledgements.} This work was supported by the Innovation Program for Quantum Science and Technology (Grant No. 2021ZD0302401), the Hunan Provincial Science Foundation for Distinguished Young Scholars (Grant No. 2021JJ10043), and the Open Research Fund from State Key Laboratory of High Performance Computing of China (HPCL) (Grant No. 201901-09).

\bibliography{Refs.bib}

\end{document}